\documentclass[twocolumn,prl,superscriptaddress,preprintnumbers]{revtex4}[10pt]
\pdfoutput=1
\usepackage{amsmath,amssymb,graphicx} %use drftcite in draftmode
\usepackage{epsf,verbatim}
\usepackage{pstricks}
\usepackage{hyperref}
\usepackage{comment}

\newcommand{\ptm}{$p_T\,$}
\newcommand{\pt}{p_T}
\newcommand{\one}{\bf 1}
\newcommand{\three}{\bf 3}

\newcommand{\mysubsection}[1]{{\bf #1~}}

\newcommand{\squishlist}{
 \begin{list}{$\bullet$}
  { \setlength{\itemsep}{0pt}
     \setlength{\parsep}{3pt}
     \setlength{\topsep}{3pt}
     \setlength{\partopsep}{0pt}
     \setlength{\leftmargin}{1.5em}
     \setlength{\labelwidth}{1em}
     \setlength{\labelsep}{0.5em} } }

\newcommand{\squishlisttwo}{
 \begin{list}{$\bullet$}
  { \setlength{\itemsep}{0pt}
     \setlength{\parsep}{0pt}
    \setlength{\topsep}{0pt}
    \setlength{\partopsep}{0pt}
    \setlength{\leftmargin}{2em}
    \setlength{\labelwidth}{1.5em}
    \setlength{\labelsep}{0.5em} } }

\newcommand{\squishend}{
  \end{list}  }

\begin{document} 

\preprint{CERN-PH-TH/2011-053}
\preprint{YITP-SB-11-06}

\title{Odd Tracks at Hadron Colliders}%

\author{Patrick Meade}
\affiliation{C. N. Yang Institute for Theoretical Physics, Stony Brook University, 
 Stony Brook, NY 11794}

\author{Michele Papucci}
\address{CERN, PH-TH, CH-1211, Geneva 23, Switzerland}
\address{on leave from Lawrence Berkeley National Laboratory, Berkeley, CA 94720}
 
\author{Tomer Volansky}
\affiliation{ Berkeley Center for Theoretical Physics, Department of Physics, University of California, Berkeley, CA 94720}
\address{Lawrence Berkeley National Laboratory, Berkeley, CA 94720}

\begin{abstract}
New physics that exhibits irregular tracks such as kinks, intermittent hits or decay in flight may easily be missed at hadron colliders.  We demonstrate this by studying viable models of light, ${\cal O}(10 \textrm{ GeV})$, colored particles that decay predominantly inside the tracker.   Such particles can be produced at staggering rates, and yet may not be identified or even triggered on at the LHC, unless specifically searched for.   In addition, the models we study provide an explanation for the original measurement of the 
anomalous charged track distribution by CDF.  The presence of irregular tracks in these models reconcile that
measurement with the subsequent reanalysis and the null results of ATLAS and CMS.   Our study clearly illustrates the need for a comprehensive study of irregular tracks at the LHC.
\end{abstract}

\maketitle

 \setcounter{equation}{0} \setcounter{footnote}{0}

%%%%%%%%%%%%%%%%%%%%%%%%%%%%%%%%%%%%%%%%%%%%%%%%%%%
\section{Introduction}\label{sec:intro}
%%%%%%%%%%%%%%%%%%%%%%%%%%%%%%%%%%%%%%%%%%%%%%%%%%%

A large variety  of new physics scenarios feature the presence of charged particles with peculiar properties, that can lead to a systematic mis-reconstruction of their tracks by the standard algorithms.  These properties can induce mismeasurements of the \ptm or even a failure to reconstruct tracks. One of the simplest examples is a particle that decays in flight inside the tracker, but in the following we list other possibilities and refer to them generically as New Odd Tracks (NOTs).  The systematic mis-reconstruction of NOTs implies that such theories may evade detection, even if they are produced at surprisingly high rates.   Consequently, particles of this kind can be very light, and may require dedicated studies for discovery. In this note we argue that there are viable models with very light colored states that would have gone unnoticed.  As a motivating example, we consider an anomaly in a recent measurement based on minimum bias events,   and provide viable explanations using NOTs.  

In~\cite{weirdcdf} the single charged particle inclusive distribution
was measured by the CDF collaboration.  The measurement  was
found to be inconsistent with the QCD prediction at high \ptm~\cite{1003.1854, 1003.2963, 1003.3433}  by
a factor of $10^4$.    Subsequent measurements by ATLAS~\cite{atlas} and CMS~\cite{cms} found no evidence for an anomaly at high \ptm.   Most recently, CDF released an erratum~\cite{errata} where they changed their track selection to remove the high-\ptm  tracks they had previously measured.  While the original result 
could 
in principle come from an experimental mismeasurement/unaccounted for background, there is an intriguing possibility that 
it was due to new physics that can also account for all subsequent findings. 

The main difference between the original CDF analysis and the subsequent reanalysis  lies in demanding a higher track quality.  Unfortunately, the reanalysis does not quantitatively find a SM explanation of the original excess tracks, which are simply removed by the quality cut.  The measurements by ATLAS and CMS also require much more stringent track quality cuts than the original CDF measurement.   To address the original anomaly, together with the null results, one is therefore required to introduce new particles which appear fundamentally different in the tracker, i.e. NOTs.    The 
more stringent cuts on the CDF data reconcile the apparent tension, however, they do so at the cost of losing the sensitivity to new physics of this kind.

The models presented are interesting in their own right.  The take-home message, however, is not the specific model.  Rather, we stress the existence of a large variety of theories that exhibit NOTs, which would be misinterpreted or even missed at the LHC unless specifically searched for.   Our work aims, in part, at motivating additional studies to ensure NOTs will not escape detection.

\section{New Odd Tracks}

Before discussing specific models, we briefly discuss the spectrum of possibilities for theories that exhibit NOTs.  It is useful to classify the possible effects of new physics  on standard track signatures.  Typically, a given model exhibits more than one signature, which may simplify its identification~\cite{simplified}:
\squishlisttwo %%\begin{itemize}
\item {\bf Kinks}.  Tracks that appear to change direction, without a secondary vertex.   Typically produced by one-prong decays.
\item {\bf Displaced vertices}.  Tracks appearing to emanate not from the PV.
\item {\bf Anomalous $dE/dx$}.  Tracks may have lower or higher ionization loss.  Standard Heavy Stable Charged Particle (HSCP) searches typically look for the latter.
\item {\bf Anomalous timing}.  Slowly moving tracks as measured via the timing module at the calorimeter, but not necessarily with a larger $dE/dx$.
\item {\bf Intermittent hits}.  Otherwise normal tracks that leave fewer hits than expected.
\item {\bf Anomalous curvature}.  Tracks that appear to bend anomalously in the tracker.
\item {\bf Stub Tracks}.  Tracks that seem to disappear inside the tracking volume.
\squishend %% \end{itemize}
We note that it is possible to misidentify some of the signatures above.   For instance, as we discuss below, tracks with kinks may be misinterpreted as  tracks with anomalous curvature.

Several of these possibilities have been explored before in the context of models of new physics.  In Gauge Mediated Supersymmetry Breaking (GMSB, see~\cite{Giudice:1998bp} and references therein), there are generically two types of NOTs.  Models with a long lived slepton as the NLSP, admit  tracks that can have kinks, anomalously high $dE/dx$ and anomalous timing.   Models with a long lived netrualino NLSP would typically give rise to displaced vertices.  Stub tracks or kinks can arise in models of Anomaly Mediated Supersymmetry Breaking (AMSB)~\cite{9904250}.  Models of SUSY with R-parity violation~\cite{rpvreview} can also give rise to kinks, displaced vertices, anomalously high $dE/dx$ and anomalous timing.  Quirks~\cite{quirks} give rise to tracks with either anomalous curvature or anomalously high $dE/dx$. 

 It is important to note that, while in many of these examples the existence of new physics can be established by other means, it could prove to be significantly harder to identify the model without studying some of the above signatures. For instance, even if supersymmetry is discovered, identifying the breaking mechanism may require the study of kinks.
There are, in addition, several interesting possibilities for NOTs that remain altogether unexplored at the Tevatron and LHC.  In particular, intermittent hits and anomalously low $dE/dx$ are possibilities that have not been investigated.  One reason for this, is that such signatures are caused by particles that leave less energy than a Minimally Ionizing Particle (MIP) in the tracking system, thereby deteriorating their reconstruction efficiency.  Similarly, models with kinked tracks may not be reconstructed, albeit having regular $dE/dx$ signature, or may present significant backgrounds from detector  material effects.

In this note we will give examples of some theories of  NOTs  that can explain the original CDF anomaly.  These models may serve as benchmarks for classes of models that will not be found through standard tracking algorithms.  A more comprehensive study of benchmark models and their prospective discovery will be presented in future work~\cite{futurework}.
For now, our hope is that the models presented here will serve, in addition to explaining the CDF  data, as motivating examples to study irregular tracking at the LHC.

\section{The CDF Anomaly}

It is useful to recall why the original CDF results are non-trivial to explain with new physics.  In~\cite{weirdcdf}, CDF looked at the \ptm distribution of all charged tracks examined through the minimum bias trigger path.  The dominant contribution to this distribution at high \ptm is the single jet inclusive channel.  The latter, however, was found to be saturated above 100 GeV by  the single charged particle distribution~\cite{Abulencia:2007ez,1003.3433}, thereby signaling the breakdown of QCD factorization.  On the other hand, the high-\ptm tracks may come from new massive particles of mass $M$.  However, on average, $\pt\lesssim M$, and therefore to account for the high-\ptm spectrum one requires particles with mass of order $100$ GeV.  In turn, the production rate for a particle of that mass is typically too low to explain the data, even if charged under QCD.  This conflict between the \ptm scale and the cross section represents the inherent difficulty in explaining this data with new physics.

Even if a scenario predicting high \ptm tracks with a large enough rate were possible, additional  constraints must be satisfied.  In particular, the new physics:
\begin{itemize}
\item Must not substantially affect the inclusive jet cross section which is well measured~\cite{Abulencia:2007ez}.
\item Can not be a new resonance that decays only into a pair of charged tracks or jets~\cite{Khachatryan:2010jd}.
\item Must not have collider-stable particles~\cite{Aaltonen:2009kea}.
\end{itemize}
With these basic restrictions the difficulty to describe the measurement with NP is understood~\cite{1003.3433}.

The tension described above can be ameliorated if the transverse momentum of new particles is mismeasured.  The presence of NOTs found in a variety of models, may thus provide an explanation to the anomaly.  A first example  which would account for a systematic mismeasurement of \ptm is a fractionally charged particle. 
Indeed, the analysis~\cite{weirdcdf} assumes that the tracks have charge one 
and therefore a particle of charge $q$ would have its \ptm measured as $p_T/q$.  A sufficiently light  new particle of this kind, interacting with QCD strength, could have a large cross section and {\em still} produce high \ptm tracks which would account for the CDF data.  

Another example of a NOT that may cause a systematic mismeasurement of \ptm is a track with a kink.  An attractive possibility    is a light mass sparticle such as a light sbottom that decays through an RPV operator.  Such a particle can be produced with a large cross-section without being detected due to the kink or the displaced vertex.  
 In a standard reconstruction algorithm these tracks {\em could} in principle, be reconstructed as a single track with a high or low \ptm and large $\chi^2$.    Only those tracks that are reconstructed with a high \ptm would rise above the background, thereby addressing the measurement.  Much like the fractionally charged particles, a model of the above kind could escape detection unless specifically searched for.

As discussed above, whether or not the original CDF data turns out to be attributed to new physics, it is important that the LHC looks for NOTs so that this window into new physics is not missed.  In fact, since the examples above cause  systematic mismeasurements in the tracker, they may not even be triggered on at the LHC.  If the CDF data is indeed a measurement of such a NOT model, the looser track quality selection criteria together with the MB triggering path, may be the only reason that these particles were observed.

\section{Light Colored Particles}
\label{sec:fractional-charges}

As discussed in the previous section, a light colored particle would have a large enough cross-section to reproduce the CDF anomaly, if the \ptm of the resulting tracks were mismeasured.    As an example, in this section we describe a viable and concrete model which exhibits fractionally charged particles.   The possibility of light sbottoms with RPV discussed above, will be presented elsewhere.

\subsection{A Model}

While fundamental particles with fractional electric charges are very constrained, composite fractionally charged particles can more easily escape detection.   For instance, let us  
 introduce
vector-like fermionic fields, $X+\bar X$, charged under the SM as
$(\three,\one)_0+(\bar\three,\one)_0$ ~\cite{Caldi:1982dj} and with a mass
$m_X\simeq\mathcal{O}(10\,\mathrm{GeV})$.   
Once produced, $X,\bar
X$ hadronize to form mesons $M_X$ and baryons $B_X$, both carrying
fractional charges.  Since the probability of hadronizing into baryons
is suppressed by an order of magnitude compared to that of mesons~\cite{pdg},
below we consider only the meson case, with charges $\pm 1/3$ and $\pm
2/3$.  
As we discuss below, if $X$, $\bar X$ were stable, they would be
excluded in Charged Massive long-lived
Particle (CHAMP) searches~\cite{Aaltonen:2009kea} by many orders of
magnitude~\cite{crap1}.  Consequently, $X$ must decay sufficiently fast and we
are therefore led to introduce additional scalar fields, $Y+\bar Y$, with quantum
numbers $(\one,\one)_{-1/9}+(\one,\one)_{1/9}$, mass $m_Y<m_X$ and
non-renormalizable couplings,
\begin{eqnarray}
  \label{eq:1}
 \frac{1}{\Lambda^2} X \bar d_R Y^3\,.
\end{eqnarray}
Here, $\bar d_R$ is the SM right handed down quark.  Different charge assignments for $Y$ can be accommodated,  implying a corresponding dimension in the operator above.  We stress, however, that some of the bounds discussed below change for different charges, and may require additional structure (such as additional fractionally charged particles).

 The virtue of
the above setup is that the colored $X$ particles are produced copiously
at hadron colliders but have suppressed production rate at $e^{+}e^{-}$ 
colliders.  On the other hand, the production rate of the fractionally charged $Y$ particles
are suppressed in both colliders due to their small EM charge.
Furthermore, as we discuss, such particles are invisible at the Tevatron since they
rarely leave ionization signals in the detectors.  Below, we study the
various constraints on this model and establish the predictions for
the Tevatron and the LHC.  

\subsection{Constraints}

New light strongly interacting particles with a fractional charge are potentially bounded by many different experiments.  Here we identify the most stringent bounds on these new states.  We find that the model above, while only marginally in some cases, evades all experimental bounds.   This is astonishing, given the lightness and strong coupling of these new states to SM particles.  Such an example demonstrates the need to carefully search for NOTs as they can easily go unnoticed, certainly in less radical scenarios.

\mysubsection{CHAMPs.}
The most straightforward way to look for new heavy states,
is by searching for slowly moving particles through time-of-flight measurements.  Such a search has been
performed by CDF~\cite{Aaltonen:2009kea} where events with isolated
muon candidates were studied as possible 
CHAMPs.  Good agreement with the SM was found,
thereby strongly excluding the possibility of a stable $X$.  For instance,
the existence of  a stable $X$ with $m_X =10$ GeV, predicts ${\cal O}(10^7)$ events that pass all
cuts, while only ${\cal O}(100)$ can be tolerated.  As a consequence, the lifetime for the $X$ decays induced by Eq.~\ref{eq:1} is strongly constrained.  For the above $X$ mass, we find the proper lifetime to be $c\tau_X \lesssim 25$~cm, corresponding to the cutoff scale, $\Lambda\simeq 3$ TeV in Eq.~\ref{eq:1}.   Interestingly, it follows that $X$ produced in colliders would typically decay inside the tracker.  

\mysubsection{Monojets.}
Since the $Y$'s do not significantly ionize, they will be registered as missing energy in events.  Consequently, one can produce a monojet by having an $X$ particle depositing much more visible energy than the other or by recoiling the  $X\bar X$ pair against a gluon. We checked the available Tevatron monojet searches~\cite{monojets}. The requirement of a jet with relatively high-\ptm, missing energy separation and track isolation rejection are sufficient to greatly suppress the number of events passing the search criteria, to a level comparable but slightly larger than their $95\%$ CL.  Since some of the cuts depend on the analysis response to the presence of the in-flight decays, one cannot asses whether the model is ruled out by this searches without a proper simulation.

\mysubsection{LEP and $e^+e^-$ colliders.}      Since $X$ couples to the SM through strong interactions, its production rate at LEP is suppressed.  The only relevant production comes through gluon radiation followed by a splitting to $X$-particles and as such is not constrained~\cite{lepstudies}.  Similarly virtual corrections to QCD observables are not constraining enough~\cite{Kaplan:2008pt}. On the other hand, $Y$-particles couple to the $Z$ boson and may therefore be constrained by the $Z$-width measurement at LEP-I.  Its small charge implies a contribution to the invisible $Z$-width, which is found to be $0.88$~MeV.    Thus even though the LEP bound is particularly strong due to a downward fluctuation, $\Gamma_{\rm inv}^{\rm NP} < 2$~MeV at $95\%$ CL~\cite{0509008}, $Y$ easily evades it.  In fact, even a $1/6$ charge is not excluded.   Finally,  constraints from lower energy  $e^+e^-$ colliders based on $dE/dx$ do not constrain the $Y$ particles either~\cite{lowenergy}.

\mysubsection{Cosmology.}   
There are no cosmological constraints on $X$ particles since they are unstable and decay almost promptly.  However, for any stable fractionally charged relic, such as the $Y$ particle, there are severe constraints on its present abundance coming from a number of searches.  The strongest comes from liquid drop experiments with mineral and silicon oil  (for a review and references see~\cite{Perl:2009zz}), requiring concentrations smaller than $O(10^{-17})$. In principle these limits do not directly translate into a relic density bound, due to large ``environmental'' uncertainties from the chemistry of the $Y$'s~\cite{Lackner:1982rq} and from additional reprocessing stages in the core of the first stars~\cite{Goldberg:1982af}, both diluting the Y's. A conservative stance would be to require a relic abundance directly compatible with such limits, $\Omega_Y h^2 \lesssim 10^{-18}$. This is clearly impossible in a standard thermal history with high reheating temperature. Lowering the reheating temperature as much as allowed by nucleosynthesis still does not help since the electromagnetic interactions are strong enough to thermally populate the Y's during reheating but not strong enough to sufficiently deplete them during the freeze-out phase. Modifying the model Lagrangian by allowing additional annihilation channels for the Y's into particles not directly coupled to the plasma lowers the abundance down to $10^{-11}\div 10^{-12}$~\cite{Giudice:2000ex}, still not sufficient to respect the bound. One possibility would be to also lower $T_{max}$, the maximum temperature  reached during the reheating era, far below $m_Y$, in order to gain a further Boltzmann suppression at the price of an extremely unnatural shape of the inflaton potential. A second, more natural, way out is to allow $Y$ to further decay to a lighter particle $Z$ with an even smaller electric charge, further relaxing the bounds from liquid drop searches.

\mysubsection{Cosmic Rays.}   
$X$ and $Y$ particles are regularly produced through Cosmic Ray (CR) interactions in the atmosphere.  A flux of $Y$ particles can then be searched for in underground experiments.  The most stringent constraint, derived by the MACRO experiment at Gran Sasso, is only sensitive to a $1/6$ charge and therefore irrelevant for the above model.  Nonetheless, it is interesting to study the bounds of $Y$ with a $1/6$ charge.
The experiment places a $90\%$ CL bound of $5\times 10^{-15}\textrm{ cm}^{-2}{\rm s}^{-1}{\rm sr}^{-1}$ on the flux at the detector, assuming an isotropic production on the rocks around it~\cite{Ambrosio:2004ub}.   Unfortunately, most $Y$ particles are produced at the top of the atmosphere and translating the above bound to a bound on the flux at the surface of the Earth is nontrivial and requires a precise modeling of the surrounding.  We compute the $Y$ flux at the detector taking into account the measured CR flux at different altitudes and considering severals models of the rocky terrain above the detector.  We find the $Y$ flux to be within a factor of two of the bound, consistent with the $90\%$ CL bound when taking systematic uncertainties into account.   A more accurate constraint could only be derived with a better study of the environment.  

Fractionally charged particles produced by CRs can also be searched for the same liquid drop experiments constraining the cosmological abundance. We compute the density of $Y$ particles accumulated at the Earth and find the model to be comfortably within current bounds.

\subsection{Predictions}
\label{sec:predictions}

The difficulty with making specific predictions for NOTs, is that it requires a detailed understanding of both the detector components and the algorithms used to reconstruct physics objects.  For instance it would be nearly impossible for us to quantify how often a moderately long lived sbottom, that decays in the tracker, would be reconstructed as a single high \ptm track, given the sizable number of tracks in the decay.  Nonetheless, for the case of a fractionally charged particle we can make a sensible set of estimates.  

\begin{figure}[t]
\begin{center}
\includegraphics[width=3.2in]{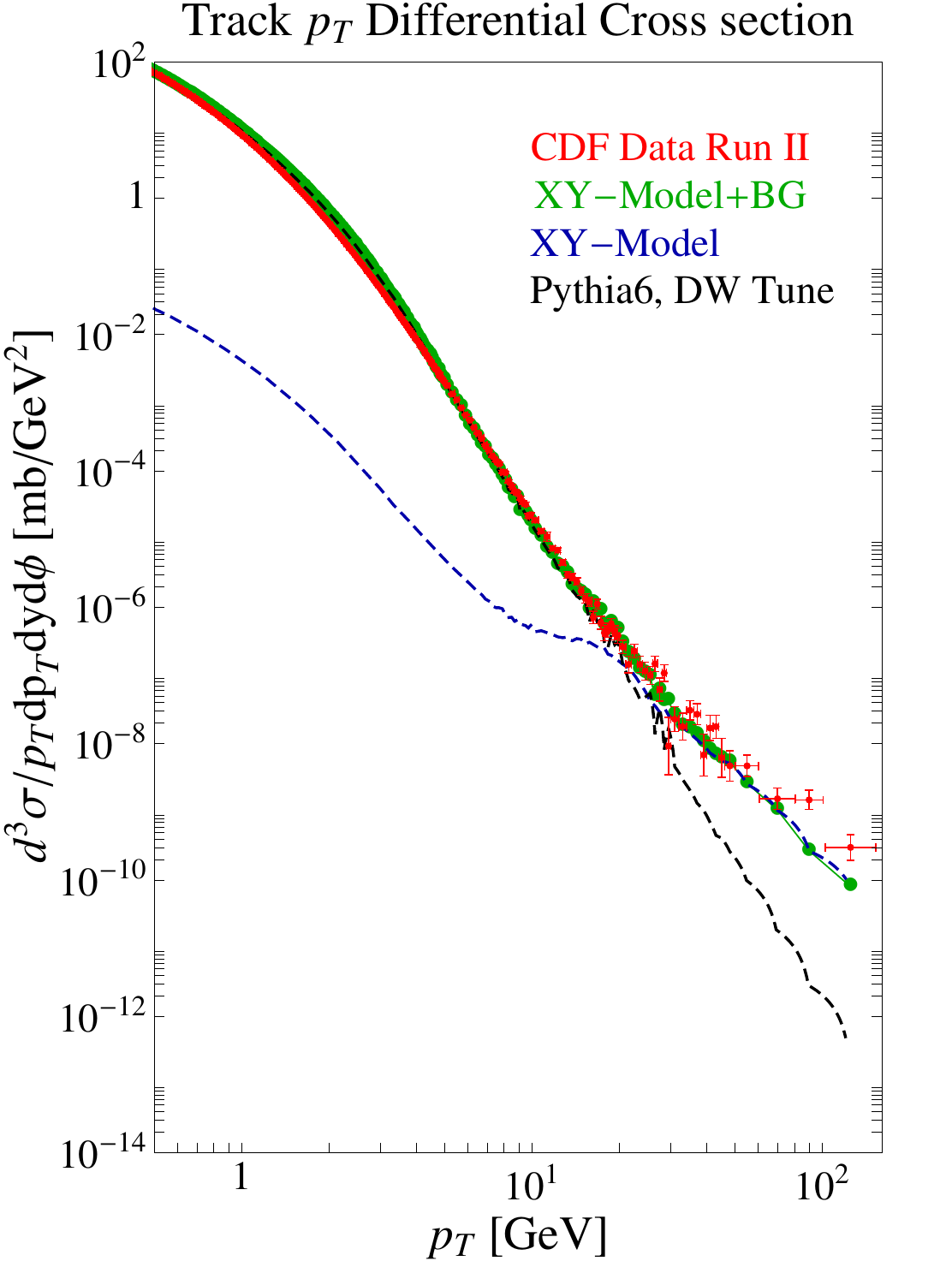}
\caption{
  Charged track $p_T$ distribution.  The dashed black line is the QCD prediction estimated using the DW tune in {\tt Pythia6.423}, while the dashed blue line is the prediction for the $X$-$Y$ model described in the text, with a best-fit value of $m_X=7$ GeV and $m_Y = 1$ GeV.  The green curve is the sum of the two contributions, to be compared with the CDF data in red.   }
\label{cdf}
\end{center}
\end{figure}

To make a prediction for CDF with the above model, we require that a measured track leaves at least 15 hits in the COT layers, and survives more than halfway through the COT before decaying.  To estimate the number of hits in the COT we use both a Landau and Bichsel parametrizations~\cite{pdg} for the tail of the energy loss of the fractionally charged mesons, and define a hit to occur when at least 15\% of a MIPs energy loss is deposited within a layer.    Using this minimal track definition, a prediction for the $p_T$ distribution of the
 $X$-$Y$ model presented above, is compared to the data in Fig.~\ref{cdf}.  Background was simulated using the DW tune in {\tt Pythia6.423}~\cite{pythia}.  

Given that $m_X$ sets the rate, adjusting $m_X$ accordingly we find a reasonable agreement with the original data published by CDF.  Based on the nature of new physics, this model predicts that at high $p_T$, tracks have fewer hits with almost no hits in the silicon tracker, and very likely a bad $\chi^2$ fit for the track. These tracks are precisely the types of tracks thrown out by CDF in their re-analysis~\cite{errata}.  It would therefore be very useful to analyze these tracks in more detail.   Of course we stress that a more accurate study of this signal is required, in order to take into account detector effects and the tracking algorithms.  Nevertheless, it clearly shows that a new physics is an intriguing and viable possibility even in light of the errata~\cite{errata}.

The cross section at ATLAS and CMS will be even higher than at the Tevatron, $\mathcal{O}(10 \mu b)$, so it would be useful to investigate how these NOTs from the model we study could show up at the LHC.  Finding kinks may be more difficult  than at the Tevatron given the larger amount of detector material in the tracker, which increases the probability of multiple scattering. To date, in order to manage backgrounds, ATLAS and CMS searches have required stringent track quality cuts.  Consequently, 
\cite{atlas,cms} would have missed NOTs of the type studied here.  Given the large production cross section, it seems advantageous to expand the current searches by loosening the track quality cuts.  In particular the nature of the silicon trackers at ATLAS and CMS allow for a lower threshold for tracker hits and may be well suited for discovering NOTs with intermittent hits.

As discussed above, while we have focused on the model parameters that could explain the original CDF data,  there is a wide range of NOT phenomenology.  In particular, the production rates may be significantly lower, thereby easing the tension with existing constraints.  Developing new techniques to search for NOTs and expanding the benchmarks beyond those given in this paper, thus provide important directions for future theoretical and experimental studies.

\section*{Acknowledgement}

We thank L.~Dixon, S.~Nussinov, M.~Reece, R.~Snider, Y.~Shadmi, G.~Sterman, T.~Phillips and  L-T.~Wang for useful discussions and G.~Elor for useful discussions and comments on the manuscript.  We especially thank N.~Moggi, M.~Mussini and F.~Rimondi for valuable discussions on the CDF measurement.  The work of PM was supported in part by NSF grant PHY-0653354.  The work of MP and TV were supported in part by the Director, Office of Science, Office of High Energy and  Nuclear Physics, of the US Department of Energy under Contract DE-AC02-05CH11231.


\begin{thebibliography}{99}
\bibitem{crap1}
There would also be severe cosmological constraints if $X$ were stable.
\end{thebibliography}
\end{document}